# Dynamics of the resistive state of a narrow superconducting channel in the ac voltage driven regime

Y. S. Yerin and V. N. Fenchenko
E-mail: yerin@ilt.kharkov.ua

*B.I. Verkin Institute for Low Temperature Physics and Engineering of the National Academy of Sciences of Ukraine, 47 Lenin Ave., Kharkov 61103, Ukraine*

## Abstract

Within the framework of the time-dependent Ginzburg–Landau equations, the dynamics of the order parameter in narrow superconducting channels of different lengths is investigated in the ac voltage-driven regime. The resistive state of the system at low frequencies of the applied voltage is characterized by the formation of time-periodic groups of oscillating phase-slip centers (PSCs). Increasing the frequency reduces the lifetime of these periodic groups. Depending on the length of the channel, the ac voltage either tends to restore the state with a single central PSC in periodic groups or minimizes the number of forming PSCs and induces their ordering in the system. For relatively short channels, a further increase in frequency leads to the suppression of the order parameter without PCS formation. For the systems with the channel length exceeding the specified limit, the formation of PSC starts after a certain time delay which increases rapidly with frequency. The current-voltage characteristics of a relatively short channel at different applied voltage frequencies are calculated. It is found that the current-voltage characteristics exhibit a step-like structure, where the height of the first step is determined by the quadruple value of the Josephson frequency.

## 1. Introduction

There exist several mechanisms through which the alternating electromagnetic field can affect the superconducting system. First of all, there is a dynamic suppression of the order parameter modulus $|\psi|$ by the effective mean square of the field amplitude $\mathbf{A}(t)$. According to the Ginzburg-Landau theory this effect is of the order $\dfrac{\delta|\psi|}{|\psi|} \sim -\dfrac{\langle \mathbf{A}^2(t) \rangle}{|\psi|^2}$.

Besides that, there is a kinetic effect of the field better known as the Eliashberg mechanism [1-3]. This phenomenon is related to the redistribution of electrons and holes in momentum space and releasing states at the Fermi level near the energy gap caused by the absorption of photons of the electromagnetic fields with the frequencies lying in the range determined by the energy gap in the spectrum and a certain parameter describing the electron energy. As a result, due to the self-consistency equation, there is an increase in the order parameter. At the macroscopic level, all of these effects manifest themselves in an increase in the critical temperature of the superconductor and the critical current. For this reason, the Eliashberg mechanism is also known as stimulation of superconductivity by an electromagnetic field [4].

Finally, the third mechanism of the effect of an alternating electromagnetic field on superconductivity is the appearance of the Shapiro steps in the current-voltage characteristics (CVCs) of superconducting systems, including the Josephson junctions [5] and, as recently found, wide superconducting films [6].

Given the fact that an alternating electromagnetic field can both suppress superconductivity and stimulate it, it is of interest to study the resistive state of a superconductor, in which the normal regions, better known as phase slip centers, lines and regions, depending on the dimensionality of the superconductor (PSC, PSL, or PSR) coexist with superconductivity in a dynamic manner within the sample. Such interest is also motivated by the fact that the Josephson junction with an ac current flowing through (ac electromagnetic field applied) exhibits complex dynamics with the emergence of a chaotic regime [7]. Since the phase slip phenomenon is, in some aspects, similar to the dissipative states in such Josephson junctions, it is logical to expect a similar behavior of the resistive state in the ac current or ac voltage-driven regime.

Experiments on the effect of an alternating electromagnetic field on the properties of superconducting systems have been first carried out in Refs. 8-11. The results of one of the first studies of this kind [10] on tin and indium whiskers have indicated the existence of "current steps" with a zero slope, the height of which is determined by the Josephson relation. In addition, the authors of Ref. 10 have observed additional steps at the subharmonics of the Josephson frequency in the hysteresis regime.

The first theoretical aimed on describing the experimental data have been done in Ref. 12. In this work, the dynamics of a PSC in the current driven regime has been studied in detail within the phenomenological formalism of Ginzburg-Landau equations for two given frequencies of the applied current. In particular, the ranges of current density where period-doubling bifurcations and chaotic solutions occur have been identified numerically.

Later it has been found that in a certain range of current density, the Poincare section of such a dynamic system behaves as a non-hyperbolic Henon map. In this case, the transition to the chaotic regime occurs according to the Feigenbaum scenario [13].

In Ref. 14, on the basis of the time-dependent Ginzburg-Landau equations, a theoretical study of the effect of time-symmetric and asymmetric alternating electromagnetic fields on the critical current of PSC formation in a quasi-one-dimensional superconducting channel in the frequency range where the Eliashberg mechanism does not work has been reported. The authors of Ref. 14 have found that at a sufficiently high power of the ac signal, the oscillations of the critical currents of the PSC formation arise. It has also been shown that in the case of an asymmetrical ac signal with zero dc component, a non-zero voltage is generated (ratchet effect).

Surprisingly, to the best our knowledge, neither theoretical nor experimental studies have been carried out to address the resistive state of a superconductor in the ac voltage driven regime. Since, as mentioned above, a quasi-one-dimensional superconducting system in the ac current driven regime exhibits a complex behavior, it is reasonable to assume that in the ac voltage driven regime, phenomena unique to this regime can be expected.

It is known that the distinctive feature of the dc voltage driven regime for a narrow superconducting channel is an S-shaped CVC, on top of which the current density fluctuations can be superimposed if the length of the system exceeds a certain value. Moreover, as has been shown in Ref. 15, for the channels of such size, there exist a voltage range, where the oscillations of the order parameter modulus are chaotic, which actually gives rise to these fluctuations. With this in mind, one of the purposes of this article is to identify the characteristic features of the ac voltage driven regime as compared to the dc voltage driven regime. In particular, it aims to elucidate the effect of the ac voltage on the evolution of the order parameter in a quasi-one-dimensional superconducting system, the processes of PSC nucleation, and find how the S-shaped CVC of the channel changes under such conditions.

## 2. Model and basic equations

The studies were carried out using the time-dependent Ginzburg-Landau equations, which, given that the system is quasi-one-dimensional, have the following form in dimensionless units:

$$u(\partial_t \psi + i\Phi \psi) - \partial_x^2 \psi - \tau \psi + |\psi|^2 \psi = 0, \quad (1)$$

$$j = -\partial_x \Phi - i(\psi^* \partial_x \psi - \psi \partial_x \psi^*). \quad (2)$$

Here, the length is measured in units of the coherence length $\xi_0 = \sqrt{\dfrac{\pi \hbar D}{8 k_B T_c}}$ at $T = 0$, where $D = \dfrac{1}{3} v_F l$ is the diffusion coefficient, and the time is expressed in units of $t_0 = \dfrac{\pi \hbar}{8 k_B T_c}$. The dimensionless order parameter $\psi$ is a complex value, which is normalized to its equilibrium value $\psi_0 = \sqrt{\dfrac{8\pi^2 k_B^2 T_c^2}{7\zeta(3)}}$ at $T = 0$, $\Phi$ denotes the electrostatic potential measured in units of $\dfrac{\hbar}{2 e t_0}$, $j$ is the density of the current flowing through the system normalized to $j_0 = \dfrac{c\Phi_0}{16\pi^2 \lambda_0^2 \xi_0}$, where $\lambda_0$ is the London penetration depth of a magnetic field at $T = 0$. A reduced temperature $\tau = 1 - \dfrac{T}{T_c}$ is introduced in the equations. We set it equal $0.1$ $(T = 0.9 T_c)$.

The numerical parameter $u$, which depends on the superconducting properties of the material and is the ratio of the relaxation time of the order parameter modulus to the relaxation time of its phase. This parameter was set equal to 1 in the calculations. Generally speaking, according to the microscopic theory, the value of $u$ depends on the amount of impurities in the superconductor, including the magnetic ones.

If $t_s T_c \ll 1$, where $t_s$ is the relaxation time on the magnetic impurities, then $u = 12$. If, on the other hand, $t_{imp} T_c \ll 1$, where $t_{imp}$ is the scattering time on impurities, the parameter $u$ takes on the value of $\dfrac{\pi^4}{14\zeta(3)} \approx 5.79$. However, the assumption of a wider range of the values of $u$ does not contradict the microscopic theory. Hence, there are no restrictions on the choice of an arbitrary positive value of this parameter.

Equations (1) and (2) are supplemented by the boundary and initial conditions corresponding to the ac voltage driven regime. Based on the selected set of normalizing parameters, these conditions are as follows:

$$\psi(0,t) = |\psi^{(0)}|, \ \psi(L,t) = |\psi^{(0)}| \exp\left(-i\frac{V_0 \sin \omega t}{\omega}\right), \tag{3}$$

$$\Phi(0,t) = 0, \ \Phi(L,t) = V_0 \cos \omega t, \tag{4}$$

$$\psi(x,0) = |\psi^{(0)}|, \ \Phi(x,0) = V_0 \frac{x}{L}. \tag{5}$$

where $L$ is the channel length, $|\psi^{(0)}| = \sqrt{\tau}$ is the equilibrium of the order parameter modulus, $V_0$ is the amplitude of the applied voltage, and $\omega$ is the signal frequency.

Equations (1) and (2) with initial and boundary conditions (3)–(5) were solved numerically using the Runge-Kutta method of the fourth order where the time and spatial derivatives were substituted by finite difference schemes. During the numerical studies of the dynamics of the order parameter modulus, time step was set equal to 0.01 and the minimum size of the spatial grid was equal to 0.5, i.e., half of the coherence length at $T = 0$.

## 3. Results and discussion

Previously, we have constructed a phase diagram that defines the length range of a quasi-one-dimensional superconducting channel and the applied dc voltage for which the states with one, two, three, or more PSCs, where the dynamics of the order parameter becomes chaotic, are realized [15].

Based on the phase diagram data, in the present work, we considered the evolution (spatiotemporal changes) of the order parameter modulus in the channels with a length corresponding to each of the above states as a function of the frequency of the applied ac voltage. In other words, we studied the systems with $L = 20$ (one PSC in the dc voltage driven regime), $L = 25$ (two PSCs in the dc voltage driven regime), $L = 35$ (three PSCs in the dc voltage driven regime) and $L = 70$ (chaotic dynamics of PSCs in the dc voltage driven regime) with the amplitude values $V_0$ taken from the range of voltages corresponding to the highest number of the PSCs for the dc voltage driven regime.

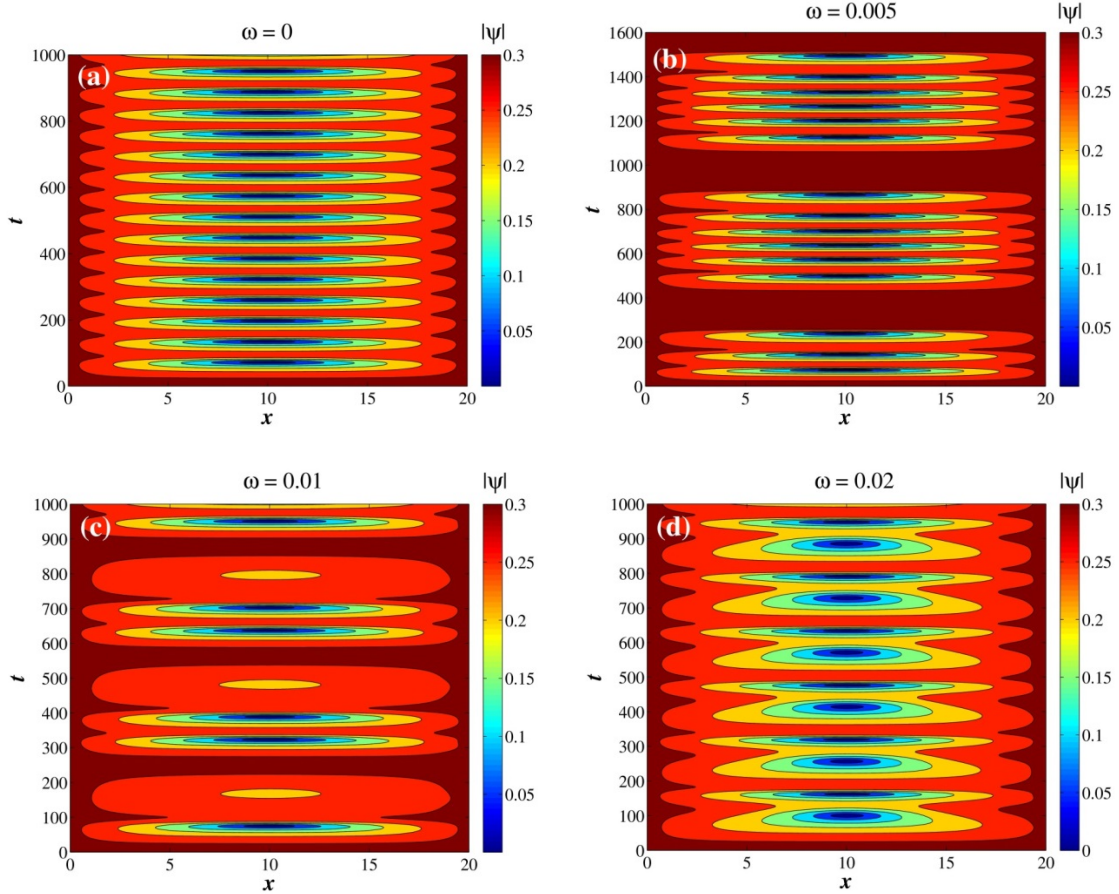

Fig. 1. Evolution (spatiotemporal dependence) of the order parameter modulus in the channel of length $L = 20$ for different frequencies of the applied voltage (the frequency values are shown above the panels). Voltage amplitude $V_0 = 0.1$. The darkest areas correspond to PSCs.

We began our investigation with the case of the channel length $L = 20$ and $V_0 = 0.1$, where only a single PSC exists in the dc voltage driven regime (Fig. 1(a)). After applying a voltage with the frequency $\omega = 0.005$ to the system, it was found that the dynamics of the central PSC can actually be represented as the groups which are regularly recurring in time, when fast oscillating PSCs periodically give way to quiet regions where the order parameter modulus is equal to its equilibrium value (Fig. 1(b)). In the following, this behavior is termed as periodic groups.

Upon doubling the frequency, periodic groups with a PCS oscillating in the center were also observed, however, as can be seen in Fig. 1(c), their lifetime decreased substantially.

Increasing the frequency further resulted in a gradual disappearance of the periodic groups and the formation of normal domains in the center of the channel (by a domain is meant a PCS increasing in size, see Fig. 1(d)). However, at a frequency exceeding about $\omega > 0.02$, as shown by numerical simulation, these domains were gradually replaced by the regions in the middle of the channel with a small but nonzero order parameter. Moreover, the higher the voltage modulation frequency, the weaker the order parameter was suppressed. For example, it was found that for $\omega = 0.021$, the minimum of the order parameter modulus was 0.1116 (approximately, a 2.7-fold decrease), while for the applied voltage frequency $\omega = 0.025$, the limiting value of the order parameter modulus was 0.2328 (a 1.29-fold decrease).

In principle, the above behavior of the resistive state of the channel is to be expected and can be explained on a qualitative level. Until the voltage is reduced to a certain level, the system has a sufficient time to create a PSC. Afterwards, in view of the signal periodicity and, therefore, decreasing voltage, the system needs more time for the formation of PSC. However, within the time required for the formation of the PSC, the voltage drops even lower, which leads to the formation of regions where the order parameter simply decreases but does not have time to vanish completely. Thus the first group of oscillating PSCs is formed in the sample. Then, the voltage rises again and the scenario repeats itself.

Increasing frequency further at fixed voltage amplitude led to the situation where PSCs have no time to form, and only the aforementioned regions are formed with the order parameter suppressed but not zero.

Let us now consider the dynamics of the resistive state in a channel of length $L = 25$ and at the amplitude $V_0 = 0.12$, which corresponds to the appearance of two PSCs in the dc voltage driven regime (Fig. 2(a)). In this regime, a characteristic feature of the system of this size is a splitting of a single PSC into two after a certain time, which depends non-monotonously on the value of $V$.

As expected, the frequency modulation at $\omega = 0.001$ generated periodic groups of the oscillating PSCs with a certain dynamics within the group, which was similar to their behavior in the voltage driven regime. In other words, as shown in Fig. 2(b), the splitting of a single PSC into two in the same way as in Fig. 2(a) (voltage driven regime) was observed in the group. However, in the end of the lifetime of those groups, the two split PSCs merge back into a single central one due to decreasing the voltage.

Increasing the frequency transformed the structure of the periodic groups (Fig. 2(c)). Their lifetime was reduced, and instead of the two splitting and reuniting PSCs, only one central PSC was realized.

Increasing the frequency further up to $\omega = 0.02$ led to the disappearance of the periodic groups and formation of a normal central domain (Fig. 2(d)).

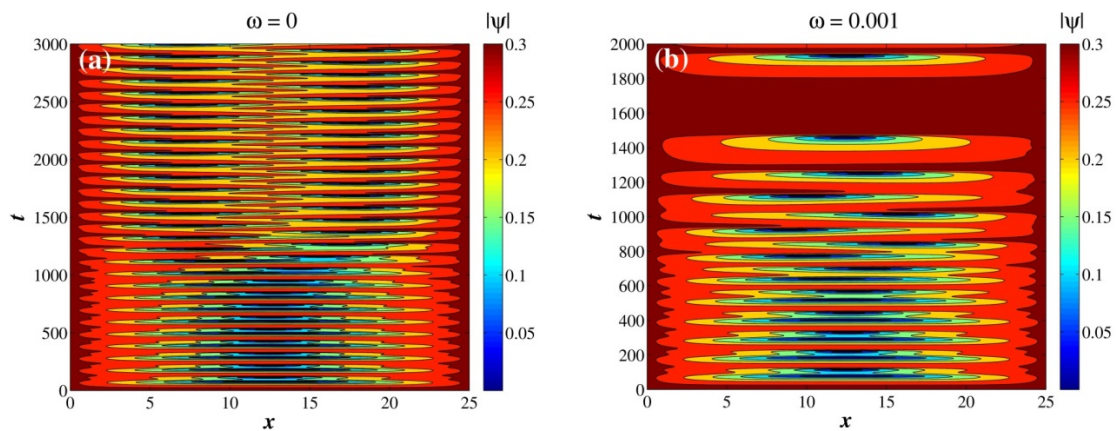

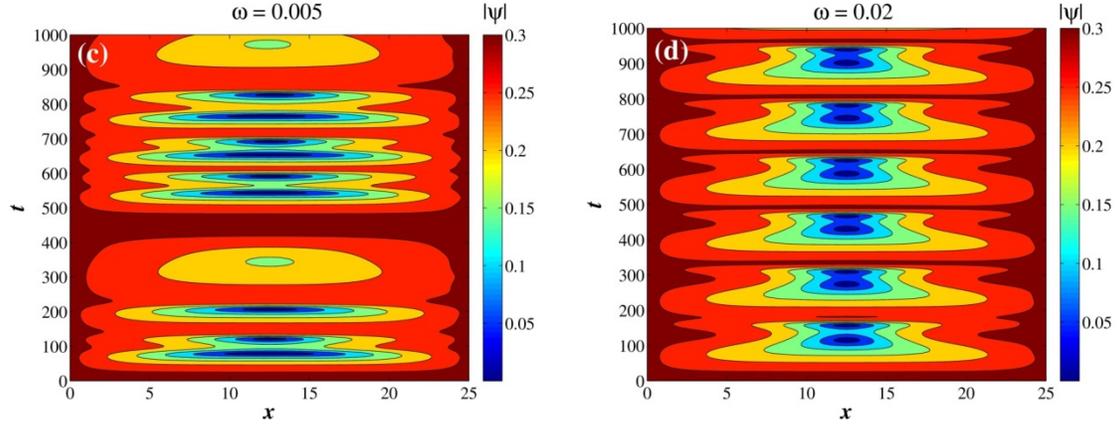

Fig. 2. Evolution (spatiotemporal dependence) of the order parameter modulus in the channel of length $L = 25$ for different frequencies of the applied voltage (the frequency values are shown above the panels). Voltage amplitude $V_0 = 0.12$. The darkest areas correspond to PSCs.

With further increase in frequency, the afore-described picture of the resistive state dynamics for a channel with $L = 20$ was observed. The order parameter was suppressed in the central region but not zero, and the degree of the suppression was inversely proportional to the signal frequency.

As shown by numerical simulation, a similar scenario, where at a frequency of approximately $\omega > 0.02$, the order parameter is suppressed without the formation of a PSC, occurs for all channels with the length not exceeding the critical value $L^* \approx 28$. It is interesting to note that $L^*$ defines the length of the system for which in the dc voltage driven regime at certain voltage values, three PSCs begin to form and kinks start to appear on the S-shaped CVC (see Ref. 15). This fact brings an added interest to the study of the influence of voltage frequency modulation on the behavior of the order parameter for the channels with $L > L^*$.

Thus, let us explore the resistive state of the system with the length $L = 35 > L^*$ and the voltage amplitude $V_0 = 0.2$, which is characterized by the formation of three PSCs (one central and two in the centers of the halves, Fig. 3(a)) in the dc voltage driven regime.

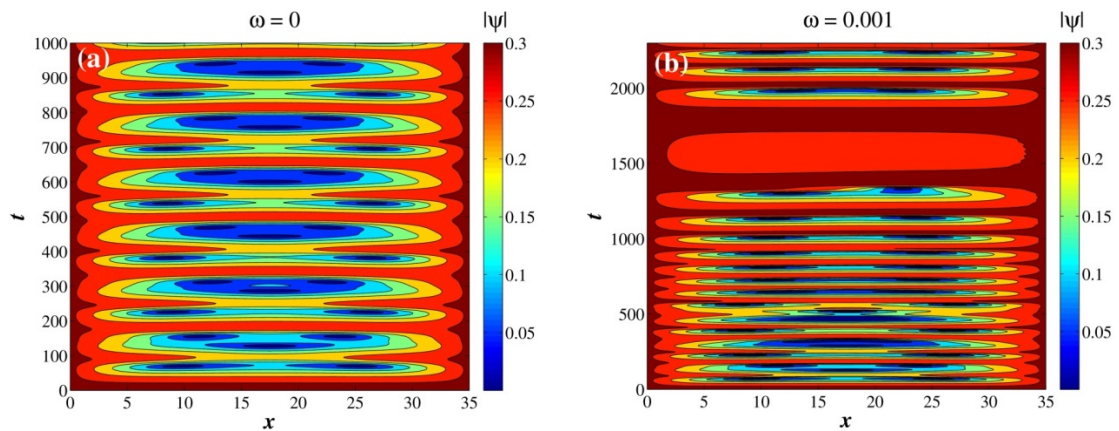

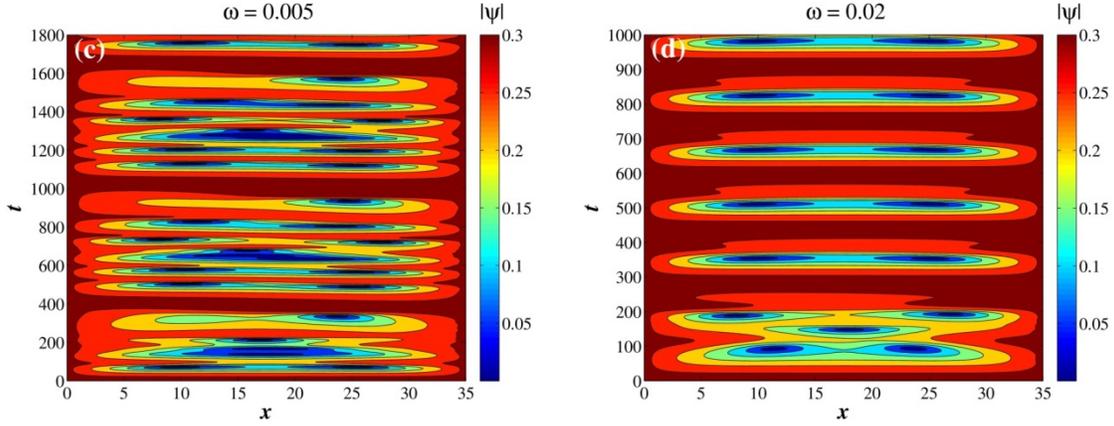

Fig. 3. Evolution (spatiotemporal dependence) of the order parameter modulus in the channel of length $L = 35$ for different frequencies of the applied voltage (the frequency values are shown above the panels). Voltage amplitude $V_0 = 0.2$. The darkest areas correspond to PSCs.

At a relatively low frequency $\omega = 0.001$ at the initial time instant, the evolution of the order parameter modulus was almost identical to its dynamics in the dc voltage driven regime. However, then the central PSC oscillating around the channel center disappeared, and only two side PSCs (located in the centers of the halves of the channel) remained in the system. This was the structure of the first group. The second group was initiated from the two side PSCs.

Voltage modulation at a frequency of $\omega = 0.005$ also predictably reduced the lifetime of the PSC periodic groups, while the actual groups were composed of two oscillating side PSCs and a single central PCS, which appeared only once in the middle of the group lifetime (Fig. 3(c)).

When increasing the frequency up to $\omega = 0.02$, it was found that the periodic groups of oscillating PSCs are transformed into a regular structure of two side PSCs (Fig. 3(d)).

In the dc voltage driven regime, the most distinctive feature of the resistive state of a quasi-one-dimensional superconducting channel with $L > L^{(\text{chaos})} \approx 48$ was the presence of the voltage range in which the order parameter behaves chaotically. As stated in Ref. 15, this behavior is caused by the formation of more than three PSCs in the channel. To clarify the effect of the ac voltage driven regime on the chaotic dynamics of the system, let us study the behavior of the channel with the length $L = 70$ and the amplitude of ac voltage $V_0 = 0.3$.

The dynamics of the order parameter at zero frequency (dc voltage) has the form shown in Fig. 4(a). The onset of modulation, as in the case of shorter channels, generated the periodic groups of oscillating PSCs with the lifetime decreasing as the frequency increases (Figs. 4(b) and 4(c)). As can be seen in Fig. 4(b), at low frequencies the periodic groups try to preserve the multiplicity of the PSCs.

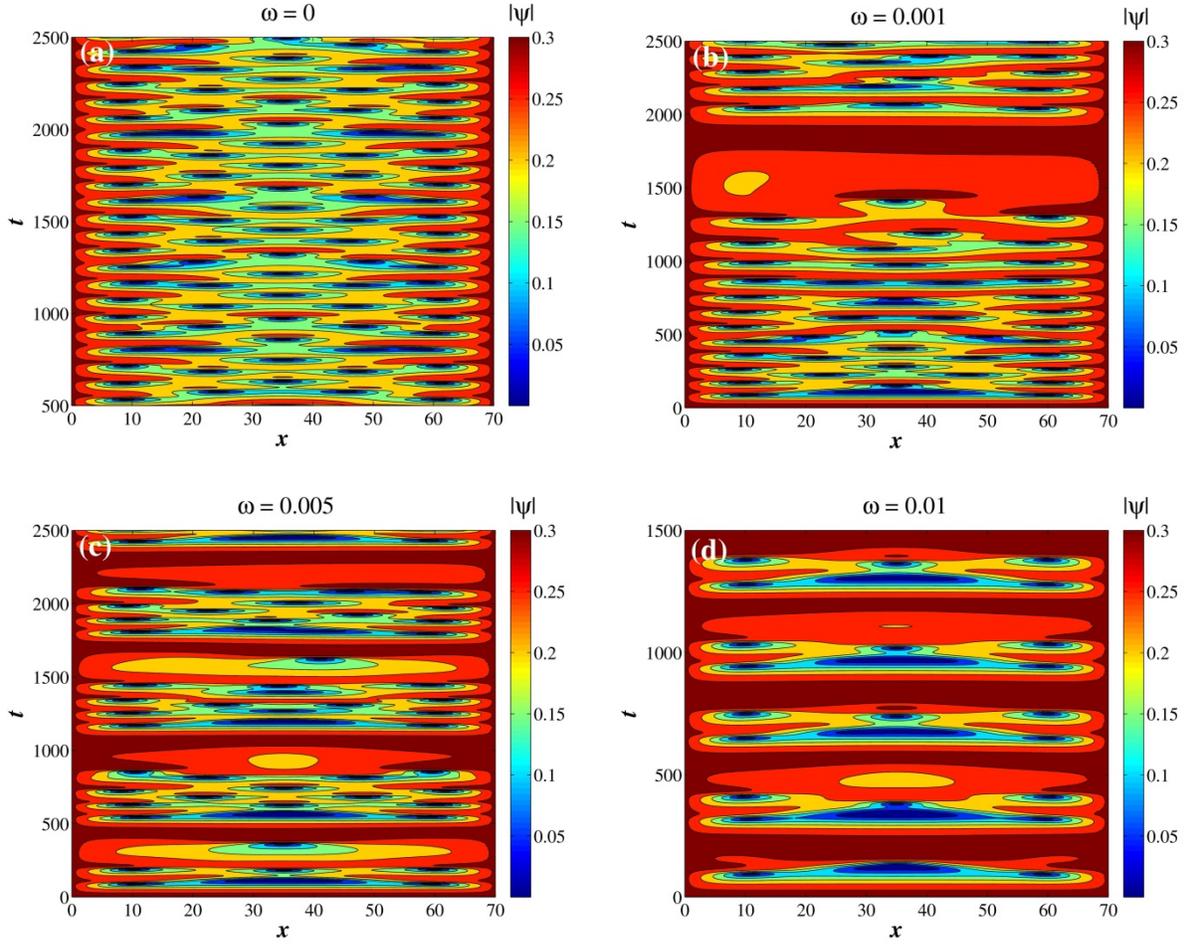

Fig. 4. Evolution (spatiotemporal dependence) of the order parameter modulus in the channel of length $L = 70$ for different frequencies of the applied voltage (the frequency values are shown above the panels). Voltage amplitude $V_0 = 0.3$. The darkest areas correspond to PSCs.

In other words, the number of oscillating PSC in the groups was still more than three. However, as we further increased the frequency, we induce ordering of the system in space and time and reduce the resistive state to regular oscillations of three PSCs (Figs. 4(a) and 4(d)).

To verify the last statement that the increase in frequency is a mechanism which induces ordering of the system, we analyzed the Fourier spectrum of the time dependence of the current density for different frequencies of the applied voltage (Fig. 5).

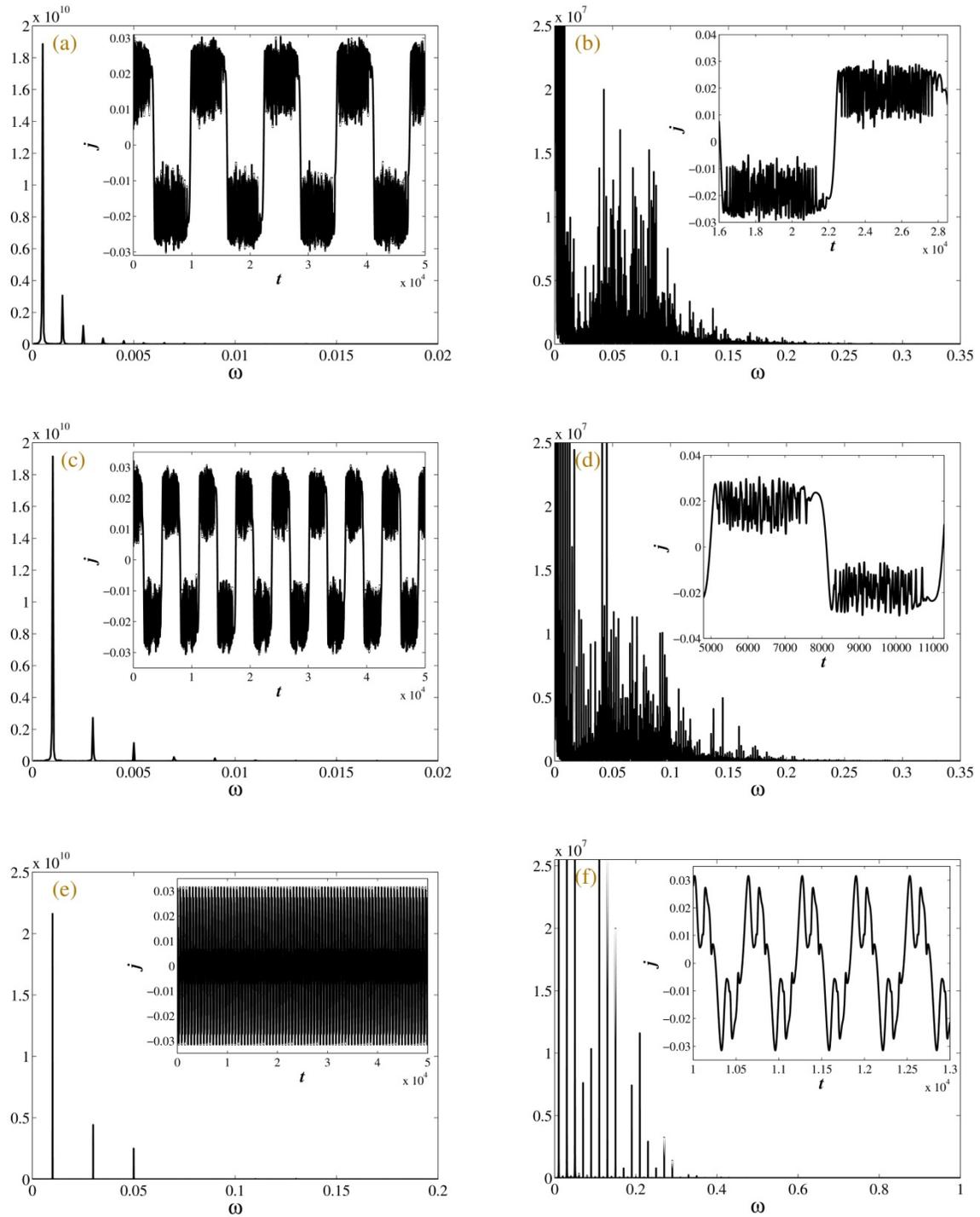

Fig. 5. Power spectra of the dependences $j(t)$ for the channel of $L = 70$ with the applied voltage amplitude $V_0 = 0.3$ and the frequencies: $\omega = 0.0005$ (**a**), $\omega = 0.001$ (**c**), and $\omega = 0.01$ (**e**). Graphs (**b**), (**d**), and (**f**) show the evolution of the spectrum $j(t)$ at higher frequencies. The insets show graphs of $j(t)$ in a wide (**a**), (**c**), (**e**) and narrow (**b**), (**d**), (**f**) time intervals.

Fig. 5(a) shows such a characteristic of a channel with the voltage modulation at the frequency $\omega = 0.0005$. The highest peak in the spectrum of the dependence $j(t)$ (this dependence is plotted in the inset in Fig. 5(a)) corresponds exactly to this frequency. The fact that the order parameter modulus exhibited a chaotic dynamics can be seen at higher frequencies (Fig. 5(b)), although the power of this signal is three orders of magnitude less than that fed to the system.

The spectrum of the time dependence of the current density for the doubled frequency has a qualitatively similar form. The main peak corresponds to the frequency $\omega = 0.001$ (Fig. 5(c)), and chaotic-like behavior of the order parameter modulus is still observed at the frequencies an order of magnitude higher than that of the signal (Fig. 5(d)), again, with the power which was three orders of magnitude lower.

However, for $\omega = 0.01$ (Fig. 5(e)), the system clearly showed signs of the ordering of the resistive state of a quasi-one-dimensional channel. The region of the spectrum which previously indicated the presence of chaotic dynamics started to show a discrete nature (Fig. 5(e)). Obviously, such discretization can be explained only by the appearance of a much more rigorous periodicity in the fluctuations of the order parameter modulus (see Fig. 4(d)).

So, let us summarize the intermediate conclusions. The resistive state of a superconducting quasi-one-dimensional channel at low frequencies of the applied voltage was characterized by the formation of the time-periodic groups of oscillating PSCs. The behavior of the PSC in these groups was similar to the analogous dynamics of the resistive state at a constant voltage. Increasing the frequency led to a reduction in the lifetime of these periodic groups. If the system in the dc voltage driven regime allows the formation of more than a single PSC, an ac voltage restored the state with one central PSC in periodic groups in the channels of length $L < L^* = 28$ and tended to minimize the number of the PSCs, if the system size exceeded this value. Increasing the frequency further to $\omega \approx 0.02$ for the systems with the length less than $L^*$, led to the suppression of the order parameter in the center of the channel. Moreover, the degree of the suppression decreased with increasing frequency. With regard to the channels longer than this length, they, as followed from the numerical modeling results, had their own specifics.

It was found that initially the regions with suppressed order parameter were formed in the channel. Then, after a certain time $T_{start}$, which increased rapidly with increasing frequency of the applied voltage (Fig. 6), an abrupt formation of PSCs occurred. The number of the PSC was determined by the length of the system (see the insets in Fig. 6).

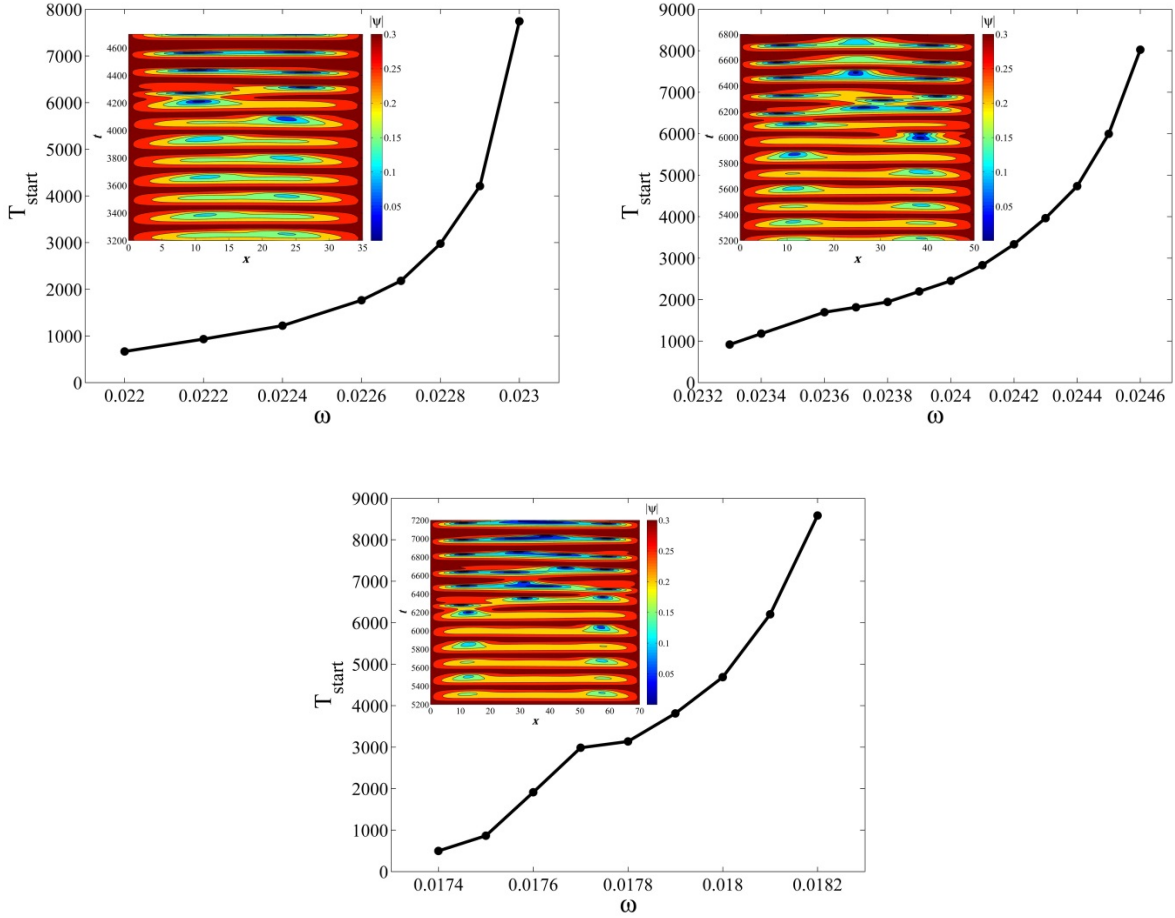

Fig. 6. Dependence of the time of PSC formation on the applied ac voltage frequency for the channels of length $L > L^*$: $L = 35$ and $V_0 = 0.2$ (top left), $L = 50$ and $V_0 = 0.3$ (top right), $L = 70$ and $V_0 = 0.3$ (bottom). The insets show the evolution of the order parameter modulus for the channels of the corresponding length and at the frequency of the applied voltage $\omega = 0.0229$ (top left), $\omega = 0.0245$ (top right), and $\omega = 0.0181$ (bottom).

The picture of the resistive state of a quasi-one-dimensional channel in the ac voltage driven regime would be incomplete without the CVC of the system. In the future, we plan to conduct a detailed study of the dependence of the CVC structure on the channel length and frequency of the applied voltage. In this paper we will only present some results for a channel with $L < L^*$ in particularly, $L = 25$.

As well known, in the dc voltage driven regime, the CVC of channels with such length begins to become S-shaped [15, 16]. We followed the evolution of the CVC of a channel with the length $L = 25$ as a function of the applied voltage frequency. The obtained numerical results are shown in Fig. 7. Note that here by CVC is meant the dependence of current density on the amplitude of the applied voltage $V_0$.

To obtain the CVC, the $j(t)$ dependence was averaged in the time interval from 500 to $5 \cdot 10^4$ with a time step of 0.01. The averaging interval was chosen such that increasing it further did not change the final value of the current density. The voltage step was 0.01.

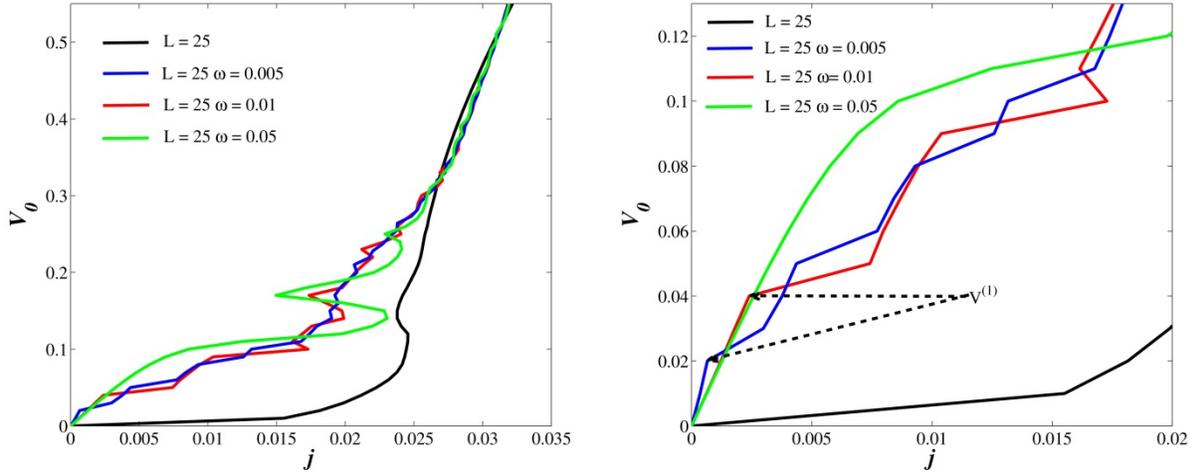

Fig. 7. CVC of the channel with $L = 25$ in the dc voltage (black curve) and ac voltage driven regimes for the frequencies indicated in the panel legend. On the right enlarged region of the CVC at low voltages illustrating the step-like structure of the dependences is shown. The dotted arrows show that the height of the first step in voltage $V^{(1)}$ is equal to the quadrupled Josephson frequency (in dimensionless units) regardless of the characteristics of the applied signal (b).

As can be seen in Fig. 7, the CVC of the channel has a step-like structure. It was found that regardless of the frequency, the value of the first voltage jump in the CVC was defined by the following equation (in dimensionless units), which is reminiscent of the Josephson relation:

$$V^{(1)} = 4\omega, \qquad (6)$$

where $\omega$ is the frequency of the applied voltage.

As shown by the analysis of the obtained data, the height of the subsequent steps followed the relation $V^{(n-3)} = n\omega$, where $n > 4$. As the frequency increased, the step-like structure of the CVC was smoothed and transformed, again, into the S-shape with a pronounced curvature.

For experimental verification of the predicted results, let us make a quantitative estimate of the frequencies and voltages using a tin channel with the length of $25\xi_0$ (5.75 µm) at $0.9T_c$. The signal frequency $\omega = 0.001$ and the amplitude $V_0 = 0.12$, at which the periodic groups of oscillating PSCs appeared (Fig. 2(b)), correspond to the linear frequency of 1.24 GHz and the voltage amplitude of 97.2 µV. The step-like structure of the CVC for the channel of such length at $\omega = 0.005$ can be experimentally observed when the voltage at a frequency of 6.2 GHz is applied. In this case, the first step in the CVC will occur at the voltage amplitude of 16.2 µV.

## 4. Conclusion

Dynamics of the order parameter for quasi-one-dimensional superconducting channels of different lengths was visualized in the ac voltage driven regime. It was found that the resistive state of the channels at low frequencies of the applied voltage is characterized by the formation of the time-periodic groups of oscillating PSCs. The behavior of PSCs in such groups was similar to the dynamics of the PSCs in the dc voltage driven regime.

Increasing the frequency of the applied voltage reduced the lifetime of the periodic groups. If the dimensions of the system are such that $L < L^* \approx 28$ and it allows the formation of more than one PSC in the dc voltage driven regime, ac voltage tended to return the channel to the state with a single central PSC in periodic groups. For the channels with $L > L^* \approx 28$, ac voltage at such frequencies minimized the amount of formed PSCs and induced their ordering in the system. In other words, ac voltage played the role of a certain "stabilizer" of the resistive state. This effect was most clearly pronounced for the channels with chaotic dynamics of the order parameter in the dc voltage driven regime.

Further increasing the frequency to $\omega \approx 0.02$ and above for the systems with the length shorter than $L^* \approx 28$ led to the suppression of the order parameter in the central part of the channel without formation of PSC. The degree of the suppression decreased with increasing the frequency. On the other hand, in the systems with the dimensions exceeding $L^*$, the area with the suppressed order parameter was also formed, however, after a certain time, which depended on the frequency of the applied voltage and the channel length, the PSC formation occured.

The CVC of the channel of length $L = 25 < L^*$ was calculated for different frequencies of the applied voltage and the stepped structure of this dependence was revealed. The height of the first step is determined by the quadrupled value of the Josephson frequency. The heights of the subsequent steps are defined by the relation $n\omega$, where $n > 4$. At higher frequencies, the step structure was smoothed and the CVC acquired a well pronounced S-shape.

In addition, using the characteristics of the superconducting state of tin, the values of frequencies and voltages for which the predicted effects can be detected experimentally in this material were calculated.

In the future, we plan to conduct a detailed study of the CVCs of longer channels and, based on the data on the evolution of the order parameter, to elucidate the origin of the step-like character of the CVCs.